\documentstyle[aps,amsfonts,amssymb,multicol]{revtex}
\input{psfig.sty}
\begin{document}
\draft
\def\vareps{\varepsilon}
\title{Unified description of Fermi and non--Fermi liquid behavior in a
conserving slave boson approximation for strongly correlated impurity models}
\author{J.~Kroha, P.~W\"olfle, and T.A.~Costi}
\address
{Universit\"at Karlsruhe, Institut f\"ur Theorie der Kondensierten
Materie, Postfach 6980, 76128 Karlsruhe, Germany}
\date{\today}
\maketitle
\begin{abstract}
We show that the presence of Fermi or non--Fermi liquid behavior 
in the SU(N)$\times$SU(M) Anderson impurity models may be read off the
infrared  threshold exponents governing the spinon and holon dynamics 
in a slave boson representation of these models. We construct a   
conserving T-matrix approximation which recovers the exact 
exponents with good numerical accuracy. Our approximation   
includes both coherent spin flip scattering and charge      
fluctuation processes. For the single--channel case the tendency to  
form bound states drastically modifies the low    
energy behavior. For the multi--channel case in the Kondo   
limit the bound state contributions are unimportant.   
\end{abstract}
\pacs{PACS numbers: 71.27+a, 71.10Fd, 75.20.Hr}
\begin{multicols}{2}
Impurity models with internal degrees of freedom and strong local 
correlations coupled to a fermionic bath have been of considerable
interest
recently\cite{nozieres.80,affleck.91,cox.88,perakis.93,cox.93}. 
The prototype is the Anderson impurity model,
involving a localized electron level (called $d$--level in the
following) hybridizing with one or several conduction
bands\cite{hewson.93}. The strong Coulomb repulsion $U$ 
($U\rightarrow \infty$) between electrons in
the localized state effectively restricts the $d$--level occupancy
to $n_d\leq 1$. The ensuing projection of Hilbert space onto the physical
subspace without multiple occupancy is a problem of 
fundamental importance in the theory of strongly correlated Fermi systems in
general. As a consequence, the Anderson model 
displays many of the salient features of strongly correlated systems,
including the formation of local magnetic moments and a competition 
between non--FL behavior caused by an incipient orthogonality
catastrophy, to which the system scales initially, and a FL fixed 
point, which is realized at energies below a characteristic scale, 
the Kondo temperature $T_K$, if the local moment can be 
completely screened by the conduction electron spin system. 
This model can, therefore,
serve as a test case for the regime of strong correlations and at the
same time for developing new methods which may later be applied
to lattice problems as well.\\
\indent
In terms of pseudofermion and slave boson operators $f_{\sigma}$, $b_m$ 
($\sigma =1,\dots , N$, $m=1,\dots , M$) \cite{barnes.76,coleman.87,read.88}
the $M$--channel Anderson model is defined by the Hamiltonian
\begin{eqnarray}
H= H_o + E_d \sum _{\sigma } f^{\dag}_{\sigma }
f_{\sigma } + V \sum _{{\vec k},\sigma ,m} 
(c^{\dag}_{{\vec k},\sigma, m}b^{\dag }_m f_{\sigma} + h.c.),
\label{H}
\end{eqnarray}
where $H_o = 
\sum _{{\vec k},\sigma ,m}(\vareps _{\vec k}-\mu )
c_{{\vec k}\sigma m}^{\dag}c_{{\vec k}\sigma m}$.
$\vareps_{\vec k}$ is the conduction band 
energy, $c^{\dag}_{{\vec k}\sigma m}$
creates a conduction electron in band $m$ with spin projection
$\sigma$ and momentum ${\vec k}$, $E_d$ denotes the energy of the 
$N$--fold degenerate local
$d$--level at position $\vec R=0$. $V$ is the hybridization matrix element
and $\mu$ the chemical potential.  
A physical electron in the local level is created by the
electron operator $d^{\dag}_{\sigma}=\sum _{m}f^{\dag}_{\sigma}b_m$, 
where the condition of no double occupancy is effected by the local
operator constraint $Q = \sum _{\sigma}
f^{\dag}_{\sigma}f_{\sigma} + \sum _{m} b^{\dag}_m b_m = 1$.
The effective coupling constant in the constrained Hilbert space is 
given by $\Gamma = \pi V^2 N(0)$, with $N(0)$ the conduction
electron  density of states at the Fermi level.
For $M>1$ the Hamiltonian Eq.~(\ref{H})
does not correspond directly to a physical system, since it involves
the existence of several empty orbital states $b_m^{\dag}|vac\rangle$,
while there is usually only one in a physical system. Introducing these
states is a mathematical convenience which allows one to derive the
SU(N)$\times$SU(M) Coqblin--Schrieffer model from Eq.~(\ref{H}) in the
Kondo limit. The latter has been studied extensively by the Bethe
ansatz method\cite{wiegmann.83}, conformal field
theory (CFT)\cite{affleck.91} and self--consistent slave boson
theory\cite{cox.93}.\\  
\indent
In this Letter we focus on the auxiliary particle Green's functions
$G_{f\sigma}(\tau_1 -\tau_2) = -\langle T \{f_{\sigma}(\tau_1) 
f_{\sigma}^{\dag}(\tau_2)\}\rangle$,  
$G_{b\ m}(\tau_1 -\tau_2) = -\langle T \{b_{m}(\tau_1) 
b_{m}^{\dag}(\tau_2)\}\rangle$.  
The angular brackets denote the statistical average in the grand
canonical ensemble,
$\langle (\dots )\rangle = 
{\rm tr} [(\dots ){\rm exp} (-\beta(H-\lambda Q))]/
{\rm tr} [{\rm exp} (-\beta(H-\lambda Q))]$. The exact projection onto the
subspace $Q=1$ is achieved by differentiating with respect to the
fugacity ${\rm exp}(-\beta\lambda)$ 
and taking the limit $\lambda \rightarrow
\infty$\cite{barnes.76,coleman.87,costi.96}. 
This procedure ensures that the projected
propagators obey \hbox{Wick's} theorem, and self--energies $\Sigma _{f,b,c}
(i\omega_n)$ may be defined by
$
G_{f,b,c} (i\omega_n)=\bigl([G_{f,b,c}^o (i\omega_n)]^{-1}-
\Sigma _{f,b,c}(i\omega_n)\bigr)^{-1},
$
where $G_{f\sigma}^o (i\omega_n)=(i\omega _n-E_d-\lambda)^{-1}$, 
$G_{bm}^o (i\omega_n)=(i\omega _n-\lambda)^{-1}$, and
$G_{cm\sigma}^o (i\omega_n)=\sum _{\vec k}(i\omega _n-\vareps
_{\vec k})^{-1}$. The $d$ electron Green's function may be expressed in
terms of $\Sigma _c$ as discussed in Ref.~\cite{costi.96}. 
The projected spectral 
functions $A_x(\omega )={\rm Im}G_x(\omega-i0)$ 
exhibit divergent threshold behavior at $\omega = 0$ with a 
proper choice of the zero of the auxiliary particle
energy\cite{costi.96}:
$A_x(\omega)\propto \omega^{-\alpha _x}$, $x=f,b$.\\ 
\indent
For the single--channel model 
with spin degeneracy $N$, which is known to have a FL
ground state, the exact exponents $\alpha _{f,b}$ have been 
determined\cite{costi.94}  
by Wilson's numerical renormalization group (NRG) approach
for the cases $N=1,2$, $M=1$.  
They may also be deduced for arbitrary $N$ by the 
following argument\cite{menge.88}: (i) In the spin screened FL 
state ($\omega , T < T_K$) 
the impurity is seen by the conduction electrons as a pure
potential scattering center. (ii) The infrared (IR) threshold behavior of
$G_{f,b}$ is then entirely due to the orthogonality catastrophe of the
overlap of the Fermi sea without the impurity level and the fully
interacting conduction electron sea affected by the potential
scattering phase shifts $\delta _{\sigma}$. The corresponding exponent
is given by $\alpha = 1 - \sum _{\sigma}(\delta _{\sigma}/\pi)^2$.
(iii) The phase shifts follow from the Friedel sum rule
$\Delta n_{\sigma}=\delta _{\sigma}/\pi$, where $\Delta n_{\sigma}$ is
the change in the number of conduction electrons at the impurity caused by
the interaction with the impurity. (iv) For the boson spectral
function the initial state is the empty impurity, which for each
spin species fills up with $\Delta n_{\sigma} = n_d/N$ conduction
electrons in the final state, until the correct impurity level
occupation $n_d$ is reached. It follows that 
\begin{equation}
\alpha _b=1-n_d^2/N.
\label{alphab}
\end{equation}
For the spectral function of fermions with spin $\sigma $  
the initial state is defined by a full impurity level with spin
$\sigma $ with the remaining $N-1$ impurity levels empty. The corresponding
change of conduction electron number in the final state with occupation
$n_d$ is $\Delta n_{\sigma} = n_d/N-1$ and $\Delta n_{\sigma '} =
n_d/N$, $\sigma '\neq \sigma$, and hence 
\begin{equation}
\alpha _f=(2n_d - n_d^2)/N.
\label{alphaf}
\end{equation}
We emphasize that the expressions Eqs.~(\ref{alphab}),(\ref{alphaf}) 
for the exponents have been confirmed using the Bethe ansatz solution and 
boundary CFT\cite{fujimoto.96}.
In the Kondo limit ($n_d=1$), $\alpha _f = 1/N$, in disagreement with
a result derived from a self--consistent parquet analysis\cite{gruneb.91}.
Note that complete spin screening is crucial for this derivation of the
exponents in terms of scattering phase shifts to be applicable: 
For the multi--channel
model ($M>1$), which exhibits a non--FL ground state, 
the exponents in the Kondo limit are known from CFT\cite{affleck.91} to be
$\alpha _f = M/(N+M)$, $\alpha _b =N/(N+M)$, while the
above argument would yield the wrong result
$\alpha _f = 1-M+(2n_d-n_d^2/M)/N$, $\alpha _b = 1-n_d^2/(NM)$.
Therefore, {\it the IR threshold exponents of the auxiliary
particles are indicators, i.e.~a necessary and sufficient condition,
for FL or  non--FL behavior, respectively}.\\
\noindent
There is evidence \cite{frota.86} that $\alpha _f$ has also observable
relevance in that it governs the 
physical electron spectral function at intermediate
frequencies $\omega \buildrel {>}\over{\sim} T_K$.\\ 
\indent
We now turn to an approximation scheme\cite{kroha.92} which is capable of 
recovering the above (exact) IR dynamics. 
As a minimal requirement, the constraint $Q=1$ has to be fulfilled
in any approximate theory. The constraint is closely related to the
invariance of the system under a simultaneous local (in time) gauge
transformation $f_\sigma(\tau) \rightarrow {\rm exp}(i\Theta
(\tau))f_\sigma(\tau)$, $b_m(\tau) \rightarrow {\rm exp}(i\Theta (\tau))
b_m(\tau)$. 
The Lagrange multiplier $\lambda$ assumes the role of a local gauge
field and transforms as $\lambda \rightarrow \lambda -
i\partial\Theta /\partial\tau $.   
Any approximation scheme respecting the gauge symmetry will preserve
the charge $Q$ in time. We shall call approximations of this type
conserving.  
Symmetry conserving approximations of the self--energies $\Sigma
_{b,f,c}$ and the irreducible vertices $\Gamma _{xy}$ may be generated
by functional
differentiation from a functional $\Phi $ of closed skeleton diagrams
as $\Sigma _x (\tau _1-\tau _2)=\delta \Phi /\delta G_x(\tau _1-\tau
_2)$, $\Gamma _{xy}=\delta ^2\Phi/(\delta G_x\delta G_y)$,
$x,y=f,b,c$.\\
\indent
We will be interested in the limit of weak
hybridization, such that the dimensionless parameter $VN(0)\ll 1$. 
Thus, let us first discuss the lowest order approximation,
which is of second order in $V$. 
The conserving approximation scheme requires the
self--energies to be determined self--consistently, which amounts to
an infinite resummation of perturbation theory even if only the second
order skeleton diagram for $\Phi$ is kept. The resulting scheme is
known as the ``Non-crossing approximation'' 
(NCA)\cite{keiter.71,kuramoto.83,mh.84}. 
It should be a 
qualitatively correct approximation, provided the perturbation series
for $\Phi$ converges, i.e.~{\it if} there are no additional collective
effects causing singularities. 
The NCA leads to very good results in the absence (i.e.~in the
multi--channel case) or sufficiently far away from a FL 
fixed point: A comparison of NCA results for the auxiliary particle
and $d$--electron spectral functions $A_f$, $A_b$, $A_d$ and exact
results obtained for the single--channel case using the NRG method
shows\cite{costi.96} that (i) the NCA auxiliary particle spectral
functions are even quantitatively correct at energies $\omega$ around
and above the Kondo scale $T_K=(M \Gamma /\pi)^{(M/N)}{\rm exp}[-\pi
|E_d|/(N\Gamma )]$, but (ii) their low energy behavior
($\omega \ll T_K$) is incorrect.  
The latter appears to be due to a lack
of vertex corrections. 
Within NCA the exponents of the above mentioned
threshold power laws may be determined
analytically as 
$\alpha_f^{NCA}=M/(N+M)$,
$\alpha_b^{NCA}=N/(N+M)$\cite{mh.84,cox.93}. For the case
$M=1$, these values disagree strongly with the exact results
discussed above.
In the multi--channel case ($M>1$), on the other hand, the NCA exponents
agree with the exponents found for the fundamental fields and their
correlation functions in CFT\cite{affleck.91} in the Kondo limit.
This suggests that the NCA describes the low energy properties
correctly in the non--FL regime of the $SU(N)\times SU(M)$
Anderson model for $n_d=1$ and that the generic behavior
of the model is that of a non--FL.\\
\indent 
It may be shown by power counting arguments that there are 
no corrections to the NCA
exponents in any finite 
order of perturbation theory\cite{cox.93}. 
However, additional collective effects, e.g.~the formation of the Kondo
singlet state, lead to FL behavior. 
Thus, it is natural 
to search for singularities in the pseudofermion--conduction
\begin{figure} 
\centerline{\psfig{figure=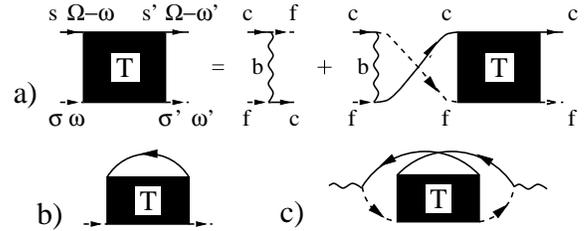,rheight=3.5cm}}
\narrowtext
\caption{a) Diagrammatic representation of the 
conduction electron--pseudofermion T--matrix $T^{(cf)}$. b)
Pseudo\-fermion and c) slave boson self--energies 
$\Sigma _{f\sigma}$, $\Sigma _b$. The terms corresponding to 
$T^{(cb)}$ are obtained by 
interchanging $f \leftrightarrow b^{\dag}$.} 
\label{tmatrix}
\end{figure}\noindent
electron scattering
channel. In particular, we consider the class of diagrams which, at
any given order of $V^2$, 
represents processes with the maximum number of spin flips.
The summation of the corresponding ladder diagrams can be
performed by solving the integral equation for the $c$--$f$  T--matrix 
(Fig.~\ref{tmatrix}a)),
\begin{eqnarray}
&T&^{(fc)}_{s\sigma,s'\sigma '}(i\omega _n, i\omega _n ', i\Omega )=
-V^2G_b(i\omega _n + i\omega _n ' - i\Omega ) \delta _{s\sigma'}\delta _{s'\sigma}
\nonumber\\
&+&V^2T\sum _{\omega _n''}G_b(i\omega _n + i\omega _n '' - i\Omega ) \times
\nonumber\\
&\ &G_{f\sigma}(i\omega _n'') G_{cs}(-i\omega _n ''+i\Omega )
T^{(cf)}_{\sigma s,s'\sigma '}(i\omega _n '', i\omega _n ', i\Omega ). 
\label{tmateq}
\end{eqnarray}\noindent
Inserting NCA Green's functions for the intermediate state 
propagators of Eq.~(\ref{tmateq}), 
we find numerically at low temperatures a pole of $T^{(cf)}$
in the singlet channel as a function of the center--of--mass (COM) frequency 
$\Omega$ in the Kondo regime ($n_d\geq 0.7$)\cite{saso.89,kroha.92}. 
This signals the tendency to form a singlet bound state 
at $\Omega =\Omega_{cf} \simeq -T_K$. 
In the empty orbital regime ($n_d\rightarrow 0$) the behavior of the
system is governed by charge fluctuations. The dominating
contributions in this low density region may be expected to result
from conduction electron--boson scattering. The corresponding 
scattering amplitude $T^{(cb)}$ is obtained from Eq.~(\ref{tmateq}) by
interchanging pseudofermions and antibosons, again leading to a pole,
at $\Omega_{cb}<0$. In the mixed valence regime
($n_d \simeq 0.5$), the poles in both $T^{(cf)}$ and $T^{(cb)}$ are of equal
importance.\\
\indent
In order to guarantee gauge invariance, self--consistency has to be
imposed. The self--energies $\Sigma _f$, $\Sigma _b$ calculated 
from $T^{(cf)}$ and $T^{(cb)}$ then follow from a 
generating functional $\Phi$\cite{kroha.92} and are depicted in
Fig.~\ref{tmatrix}b),c). 
They are given as nonlinear and nonlocal (in time)
functionals of the Green's functions. The Green's functions in turn
are expressed in terms of the self--energies, closing the set of 
self--consistent equations (conserving $T$--matrix approximation, CTMA). 
Note that the contribution to $\Phi$ containing one boson rung 
corresponds to NCA. 
The diagram with two rungs is excluded since it is not a skeleton.
The sum of the $\Phi$ diagrams with up to four rungs constitutes a large
$N$ expansion correct up to $O(1/N^2)$ and is identical to the
diagram class used in Ref.~\cite{anders.94}. We emphasize that the \hbox{CTMA},
i.e.~the  {\it selfconsistent} summation of the {\it infinite} series of all diagrams
shown  in Fig.~{\ref{tmatrix}} is justified on physical as well as formal grounds: At
any loop order  of $\Phi$ it includes (1) the maximum number
of spin flip as well as charge fluctuation processes; (2) all leading and 
sub--leading IR singular contributions, because all terms not
included cancel pairwise in the IR regime
\cite{kroha.97}.                                                                                                         
The threshold property of the auxiliary spectral functions implies 
that the exact T--matrices 
$T^{(cf)}$ and $T^{(cb)}$ have no spectral weight at negative COM frequencies
$\Omega $, in contrast to the poles 
appearing in the ``perturbative'' evaluation, i.e.~inserting NCA 
propagators as discussed after Eq.~(\ref{tmateq}).
Consequently, these poles are shifted to $\Omega =0$ by self--consistency,
where they merge with the continuous spectral weight present for $\Omega >0$,
thus 
renormalizing the threshold exponents of the auxiliary 
\begin{figure} 
\centerline{\psfig{figure=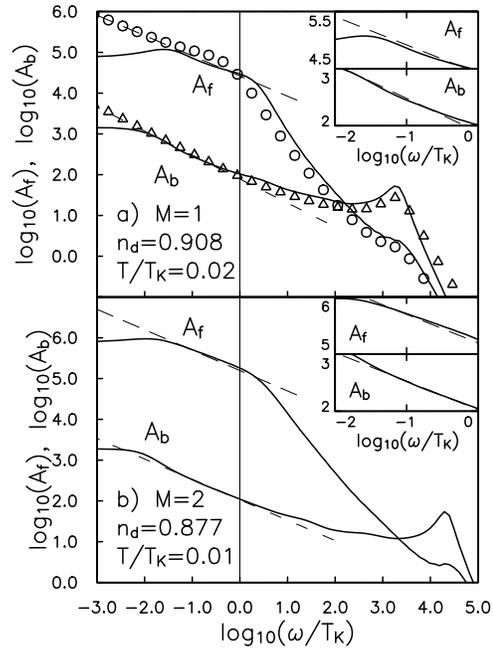,rheight=9.0cm}}
\caption{Pseudofermion and slave boson spectral functions $A_f$ and $A_b$
in the Kondo regime ($N=2$; 
$E_d=-0.05$, $\Gamma =0.01$ in units of the half--bandwidth), 
for a) the single--channel ($M=1$) and b) the
multi--channel ($M=2$) case.
In a) the symbols represent the results of NRG for the same parameter set,
$T=0$. The slopes of the dashed lines indicate the exact
threshold exponents as given by Eqs.~(2),(3) for $M=1$ and by CFT 
for $M=2$. Deviations from the power laws at low frequencies $\omega$ 
shift towards $\omega =0$ as $T\rightarrow 0$, i.e.~are finite $T$
effects. Insets show magnified power law regions.} 
\label{spectra}
\end{figure} \noindent
spectral functions, as seen below. This is an expression of the fact 
that the Kondo singlet is not
a two--particle bound state but rather  a collective many--particle state.\\
\indent
After analytical continuation to the real
frequency axis we have solved the CTMA numerically by
iteration. In the Kondo regime ($n_d\geq 0.7$) of the $N=2$, $M=1$ model, 
we have obtained reliable results down to temperatures 
of the order of $10^{-2}T_K$ (Note that $T_K\rightarrow 0$ in the Kondo    
limit). In the mixed valence and empty impurity 
regimes, significantly lower temperatures may be reached, compared to the 
low temperature scale of the model. 
As is shown in Fig.~\ref{spectra}a), the
spectral functions obtained are in good agreement with the results of NRG
(zero temperature results), given the uncertainties in the NRG at higher
frequencies. Typical behavior in the Kondo regime (Fig.~\ref{spectra}a))
is recovered: a broadened peak in $A_b$ at $\omega\simeq |E_d|$, representing
the hybridizing $d$--level and a structure in $A_f$ at 
$\omega \simeq T_K$. Both functions display power law behavior at frequencies
below $T_K$, which at finite $T$ is cut off at the scale $\omega
\simeq T$. The exponents extracted from the frequency range
$T<\omega<T_K$ of our finite $T$ results  
compare well with the exact result also shown
(see insets of Fig.~\ref{spectra}). A similar analysis has been 
performed for a number of parameter sets spanning the 
\begin{figure} 
\centerline{\psfig{figure=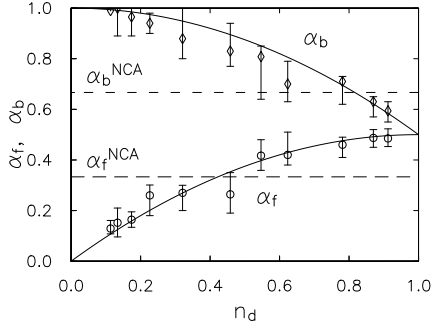,rheight=4.5cm}}
\caption{CTMA results (symbols with error bars) for the threshold 
exponents $\alpha _f$ and $\alpha _b$ of $A_f$ and $A_b$, $N=2$, $M=1$. 
Solid lines: exact (Eqs.~(2),(3)), dashed lines: NCA results.}
\label{exponents}
\end{figure}\noindent 
complete range of
$d$--level occupation numbers $n_d$. 
The extracted power law exponents are shown in Fig.~\ref{exponents}, 
together with error bars estimated 
from the finite frequency ranges over which
the fit was made.
The comparatively large error bars
in the mixed valence regime arise because  
here spin flip and charge fluctuation processes are of equal importance, 
inhibiting the convergence of the numerical
procedure. In this light, the agreement
with the exact results is very good. This is evidence that 
{\it the CTMA recovers the signature of FL behavior 
present in the exact auxiliary particle dynamics
of the single--channel Anderson model}.
As a consequence of the conserving scheme, the FL behavior
should be reproduced in physical properties as well,
when derived from the same generating functional $\Phi$. These
evaluations are in progress.\\
\indent
In the multi--channel case with $M>1$, $N=M$, it follows from the 
symmetry of the model under the transformation 
$f_{\sigma}\rightarrow b^{\dag}_{\sigma}$,
$b_{m}\rightarrow f^{\dag}_{m}$, $E_d\rightarrow -E_d$
that $\alpha _{f,b}(n_d )=\alpha _{b,f}(1-n_d)$. For $n_d=1$, 
$\alpha _{f,b}$ are known from CFT (see above). It follows that NCA yields the
exact exponents both in the Kondo and in the empty impurity limits of
the multichannel model. At present, it is not known whether this is
the case for arbitrary $n_d$. 
Using NCA Green's functions as discussed above, 
we find again a pole in the spin singlet channel of the
$T$--matrix of pseudo\-fermions ($f_s, s=\uparrow,\downarrow$) and
conduction electrons of flavor $m$ ($c_{\sigma m}$) for $n_d < 1$.
However the weight of the pole vanishes 
for $n_d\rightarrow 1$, and a numerical solution of the 
self--consistent CTMA equations for $M=N=2$, $n_d=0.877$ 
indeed yields exponents $\alpha _f\simeq 0.44$,
$\alpha _b\simeq 0.49$, very close to the exact value of $1/2$ for $n_d=1$
(Fig.~\ref{spectra}b)). This is consistent with the fact that there
should not be a bound state contribution in the overscreened case.\\ 
\indent
In this Letter we have considered the dynamics of the auxiliary particles
for the (N,M) generalized Anderson impurity model with particular
emphasis on the realization of FL behavior.  
We have shown that the occurrence of FL behavior can be deduced
from the IR threshold exponents of the auxiliary particle
spectral functions. A conserving self--consistent approximation incorporating
an infinite number of coherent spin flip and charge transfer processes 
(CTMA) leads to singular contributions
which renormalize the threshold exponents by self--consistency. 
There cannot be a renormalization of the exponents in 
any finite order self--consistent summation.
A numerical evaluation of the CTMA yields good agreement with the
known exact values in the single--channel case, indicating that CTMA 
recovers the FL behavior. By contrast, in the multi--channel case
the singular contributions are ineffective in the limit $n_d\rightarrow 1$
of the two--channel Anderson model, i.e.~the 
non--FL state persists and the exponents known from CFT are recovered.\\
\indent 
We are grateful to P.~J.~Hirschfeld and K.~A.~Muttalib for stimulating
discussions. This work was supported in part by DFG.
Part of the work by J.K. was performed at LASSP, 
Cornell University, supported by the A.~v.~Humboldt Foundation and
NSF grant no.~DMR-9407245.


\newpage
\end{multicols}

\end{document}